\documentclass[prb,twocolumn,superscriptaddress,amsfont,amssymb,amsmath,showpacs,balancelastpage,nofootinbib]{revtex4-2}
\usepackage[utf8]{inputenc}
\usepackage{CJKutf8}

\usepackage{natbib}

\usepackage{amsmath,amssymb,amsfonts,tabu}
\usepackage{physics}
\usepackage{dsfont}
\usepackage{graphicx}
\usepackage{comment}

\usepackage[normalem]{ulem}

\usepackage{cancel}
\usepackage{subcaption}
\usepackage{tikz}
\definecolor{cof}{RGB}{219,144,71}
\definecolor{pur}{RGB}{186,146,162}
\definecolor{greeo}{RGB}{91,173,69}
\definecolor{greet}{RGB}{52,111,72}

\usetikzlibrary{arrows,shapes,trees} 
\usetikzlibrary{positioning,calc}
\usetikzlibrary{decorations.markings}

\newcommand{\cD}{\mathcal D}
\newcommand{\cE}{\mathcal E}

\newcommand{\Z}{{\mathbb Z}}
\def\U{\mathrm{U}}

\usepackage[left=1in,right=1in,top=1in,bottom=1in]{geometry}

\usepackage{hyperref}

\usepackage{tikz-cd}

\hypersetup{colorlinks=true,urlcolor=[rgb]{0,0,0.5},citecolor=[rgb]{0,0.5,0},linkcolor=[rgb]{0,0.5,0}}

\makeatletter
\def\l@subsubsection#1#2{}
\makeatother

\begin{document}

\title{Detection of 2D SPT phases under decoherence}

\author{\large Naren Manjunath}
\thanks{These authors contributed equally to the work}
\email{nmanjunath@perimeterinstitute.ca}
\affiliation{\footnotesize Perimeter Institute for Theoretical Physics, 31 Caroline St N, Waterloo, ON N2L-2Y5, Canada}

\author{\large Alex Turzillo\footnotemark[1]\footnotemark[3]}
\thanks{These authors contributed equally to the work}
\email{aturzillo@perimeterinstitute.ca}
\affiliation{\footnotesize Perimeter Institute for Theoretical Physics, 31 Caroline St N, Waterloo, ON N2L-2Y5, Canada}

\author{\large Chong Wang}
\affiliation{\footnotesize Perimeter Institute for Theoretical Physics, 31 Caroline St N, Waterloo, ON N2L-2Y5, Canada}

\begin{abstract}
    We propose a bulk order parameter for extracting symmetry-protected topological (SPT) invariants of quantum many-body mixed states on a two dimensional lattice using partial symmetries. The procedure builds on the partial symmetry order parameter recently developed by some of the authors to study SPT phases of pure states and adapts them to the decohered setting. For a symmetry $G = E \times A$ where $E$ is a strong symmetry and $A$ is a weak symmetry, we show that the partial symmetry order parameter detects SPT invariants jointly protected by $E$ and $A$. We demonstrate this explicitly using a class of mixed states obtained from CZX-type models with $\Z_2\times\Z_2$ symmetry and subjecting them to noise that weakens one of the $\Z_2$ symmetries. We also comment on the practical detection of SPT invariants in quantum simulators through randomized measurements.

\end{abstract}

\maketitle

The program of classifying and characterizing mixed-state phases of matter has progressed rapidly in recent years \cite{coser2019classification,Buca_2012,me_open1,MaWangASPT,ma2023topological,Ma2024SPTdoubled,XueLeeBao,Lee2025MixedSPT,LessaChengWang,Wang2025AnomalyOQS,XuJian2025,lu2024bilayer,Guo2025LPDO,ZhangStrange,Chen2024separability,Su2024CFT-CI,Guo2024MixedSPT,kiely2025dephasing}. In particular, there are proposals to mathematically classify symmetry-protected topological (SPT) phases under decoherence, using decorated domain wall constructions \cite{MaWangASPT,ma2023topological}, the doubled Hilbert space formalism \cite{Lee2025MixedSPT,Ma2024SPTdoubled,XueLeeBao}, boundary anomalies \cite{LessaChengWang,Wang2025AnomalyOQS,XuJian2025}, canonical purification \cite{lu2024bilayer} and local purification \cite{Guo2025LPDO}. New mixed state order parameters, including generalized string order parameters \cite{me_open1}, strange correlators \cite{ZhangStrange,Lee2025MixedSPT} and separability measures \cite{Chen2024separability}, have elucidated the effects of decoherence on topological phases in one and two space dimensions \cite{Chen2024separability,Su2024CFT-CI,Guo2024MixedSPT,kiely2025dephasing}. However, the existing theory has significant limitations in (2+1) dimensions. Here only a partial set of bulk SPT order parameters is known even for pure states with (finite, 0-form, internal) Abelian group symmetry (see \cite{PerezGarcia2008,Tantivasadakarn_2017,turzillo2025SPT} for the progress till date). Moreover, while the boundary anomalies of mixed-state SPTs have been studied, we currently do not know of order parameters to detect them in the \textit{bulk}.

In this paper, we consider (2+1)-dimensional many-body mixed states with symmetry $G = E \times A$, where $E$ is a strong symmetry and $A$ is a weak symmetry\footnote{The classification of SPT phases becomes richer when the strong and weak symmetries do not form a simple direct product \cite{ma2023topological}. Our methods and results in this work can be extended to the more general case, but for technical simplicity we shall not describe this in detail here.}. Our working example is a class of exactly solvable models for (2+1)-dimensional pure state bosonic SPTs with $G = \Z_2\times \Z_2$ symmetry, generalizing the well-known square lattice CZX model \cite{Chen_2011}. We evaluate its SPT invariants using a \textit{partial symmetry order parameter} denoted $I_{g,M}$, recently developed in \cite{turzillo2025SPT} for systems with a $C_M$ rotation symmetry. For the case of $\Z_2^2$ symmetry, this bulk order parameter in fact gives a complete characterization. We then consider a mixed-state version in which $E = A = \Z_2$, meaning that one of the $\Z_2$ subgroups breaks down to a weak symmetry. We propose a mixed-state SPT order parameter denoted $J_{g,h,M}$, which generalizes $I_{g,M}$. A general result, which we explicitly verify for the CZX example, is that the SPT indices protected \textit{only} by the weak $A$ subgroup are no longer well-defined. However, \textit{all} other SPT invariants of the $G$-symmetric pure state are detected by combinations of $J_{g,h,M}$. Invariants that depend only on the strong symmetry can be detected by letting $h$ be trivial, and reduce to $I_{g,M}$, whereas those \emph{jointly} protected by the strong and weak symmetries require both $g$ and $h$ to be non-trivial, and reduce to $I_{gh,M}I_{h,M}^{-1}$ for pure states. This result concretely verifies the prediction \cite{MaWangASPT,ma2023topological,Lee2025MixedSPT,LessaChengWang,Wang2025AnomalyOQS,Ma2024SPTdoubled} that mixed-state bosonic SPT phases in $d$ space dimensions are classified by the group $\mathcal{H}^{d+1}(G,\U(1))/\mathcal{H}^{d+1}(A,\U(1))$. $J_{g,h,M}$ detects mixed-state SPT invariants to the same extent that $I_{g,M}$ distinguishes pure state SPT invariants: this depends on the available degrees of rotation $M$ \cite{turzillo2025SPT}. 
We conclude by discussing how $J_{g,h,M}$ should in principle be measurable using a Hadamard test or with randomized measurements. 

\textit{Pure state SPT calculation.} Let $G = \Z_2^2$, with a general group element written as $g = (g_a, g_b)$. The $\Z_2^2$ SPT phases are classified by the cohomology group $\mathcal{H}^3(G,\U(1)) = \Z_2^3$ \cite{Chen2013}. Taking $i, j \in \{a,b\}$, a representative cocycle in this group is \cite{Tantivasadakarn_2017}
\begin{equation}\label{eq:omega-def}
    \omega = n_a \omega_a + n_b \omega_b + n_{ab} \omega_{ab}
\end{equation}
where
\begin{align}\label{eq:omegaij-def}
    \omega_i(g,h,k)=(-1)^{g_ih_ik_i}; ~
    \omega_{ij}(g,h,k)&=(-1)^{g_ih_ik_j}.
\end{align}
The SPT invariants are $n_a,n_b,n_{ab}$. The basis cocycles $\omega_i, \omega_{ij}$ are referred to as Type-I and Type-II respectively, and $\omega$ is said to be Type-I or II if it has components of Type \emph{at most} I or II.\footnote{Under a change of generators of $\Z_2^2$, the $\omega_i$ map to sums of $\omega_i$, whereas $\omega_{ab}$ becomes a sum of the $\omega_i$ and $\omega_{ab}$. Thus, the Type of $\omega$ is an generator-independent notion.} The invariant $\omega_{ab}$ is said to be \emph{jointly protected} by the $\Z_2$ factors $a$ and $b$. In general we also need to consider `Type-III' cocycles, which are protected by a product of three symmetries; we will comment on this case later.

\begin{figure}[t]
    \centering
    \includegraphics[width=0.7\linewidth]{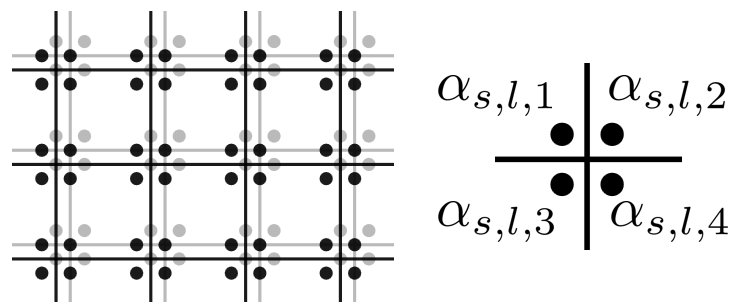}
    \caption{Notations for the two-layer CZX model.}
    \label{fig:CZX}
\end{figure}

We will now write a lattice model that realizes each choice of $(n_a,n_b,n_{ab})$. Consider two layers $l = a, b$ of a square lattice on a torus or infinite plane. At each site $s$ there are two vertices $(s,l), l=a,b$. Each vertex has four $\Z_2$ valued degrees of freedom $\alpha_{s,l,j} \in \{0,1\}, j=1,2,3,4$, as shown in Fig.~\ref{fig:CZX}. When the specific choice of $s$ is clear, we will just write $\alpha_{l,j}$.

Consider the eight-spin state on a single site, denoted $\ket{\{\alpha_{s,l,j}\}}$. The CZX state is a superposition over all `plaquette-GHZ' configurations, in which $\alpha_{s,l,j} = \alpha_{s',l,j'} = \alpha_{l,p}$ whenever $(s,j),(s',j')$ belong to the same plaquette $p$:
\begin{equation}
    \ket{\Psi} := \bigotimes_p \sum_{\alpha_{a,p},\alpha_{b,p}} \ket{\alpha_{a,p},\alpha_{b,p} }.
\end{equation}

The topological invariant appears in the definition of the symmetry operators $\hat{U}_g = \otimes_s \hat{U}_{g,s}, g 
\in \Z_2^2$. We define the operators $X_{s,l,j}$, which takes $\ket{\alpha_{s,l,j}} \rightarrow \ket{\alpha_{s,l,j}+1}$, and $CZ_{s,l,j_1,j_2}$ which takes $\ket{\{\alpha_{s,l,j}\}} \rightarrow (-1)^{\alpha_{s,l,j_1}\alpha_{s,l,j_2}} \ket{\{\alpha_{s,l,j}\}}$. This is a controlled-$Z$ operator acting on a single layer $l$. Let $X_{s,l} = \otimes_j X_{s,l,j}$ and $CZ_{s,l} = \otimes_j CZ_{s,l,j,j+1}$. 
Now the single-site symmetry action corresponding to $\omega$ in Eq.~\eqref{eq:omega-def} can be implemented as follows (below we have dropped $s$, and retained only the $l$ subscript):\footnote{Like the single-layer CZX model, these models can be obtained from the general group cohomology construction \cite{Chen2013} up to some local $Z$ gates in the symmetry action, which do not affect the topology.}
\begin{align}\label{eq:PureCZX-SymAction}
    U_g &= CZ_a^{g_an_a}CZ_b^{g_bn_b}CZ_a^{g_bn_{ab}}  X_a^{g_a}X_b^{g_b}. 
\end{align}
The $X$ action flips each $\alpha_{l,j}$ when $g_l \neq 0$. The $CZ$ operators give an additional phase which depends on $n_a,n_b,n_{ab}$.
For instance, suppose only $n_a \neq 0$, and choose $g=g_a$. Then $U_g = CZ_a X_a$, and we get a phase $(-1)^{\sum_j \alpha_{a,j}\alpha_{a,j+1}}$. If only $n_{ab} \neq 0$, applying $g_b$ gives this same phase (note that it depends on the $a$ variables). Finally, the system has $C_4$ rotation symmetry about either a vertex or a plaquette, which takes $(s,j) \rightarrow (s',j') = {^{C_4}}(s,j)$ without any extra phase factors.

Next we describe how to detect these SPT invariants, following~\cite{turzillo2025SPT}. Let $g \in G$ be a general group element with symmetry operator $U_g$. Suppose the system has an $M$-fold rotation symmetry generated by $C_M$. Assume $g^M = e$, the identity. We consider a $C_M$-symmetric region $\cD$, and define
\begin{equation}\label{eq:IgM-def}
    I_{g,M} := \operatorname{arg} \bra{\Psi}U_g^{\cD} C_M^{\cD} \ket{\Psi}. 
\end{equation}
$\mathcal{O}^{\cD}$ denotes the restriction of the operator $\mathcal{O}$ to $\cD$. Ref.~\cite{turzillo2025SPT} evaluated $I_{g,M}$ for the general class of fixed point Chen-Gu-Wen group cohomology models \cite{Chen2013} and checked that the result is coboundary-invariant. Some evidence for its quantization away from fixed points is available through a combination of conformal field theory arguments \cite{Shiozaki2017point,zhang2023complete,kobayashi2024FCI} and numerics \cite{Shiozaki2017point,zhang2023complete,kobayashi2024FCI,turzillo2025SPT}. The partial rotation order parameter has previously been used to characterize crystalline invariants in integer and fractional Chern insulators \cite{Shiozaki2017point,zhang2023complete,kobayashi2024FCI}, as well as higher central charges in topologically ordered states \cite{Kobayashi2024HigherCC}. Indeed, in the case of SPT states $I_{g,M}$ generally measures a combination of internal and crystalline SPT invariants. Ref.~\cite{turzillo2025SPT} showed that on a square lattice, with $g^2 = e$, one can isolate the internal SPT invariant by defining
\begin{equation}\label{eq:Ig-def}
    I_g := I_{g,4} I_{g,2} I_{e,4}.
\end{equation}
This formula satisfies some important checks: (i) it gives the same result whether we use a vertex or a plaquette center as our origin, and (ii) $I_g$ is independent of the size and shape of the region $\cD$, although the magnitude of the expectation value in Eq.~\eqref{eq:IgM-def} decreases with $|\partial \cD|$. We can make a prediction for $I_g$ evaluated on an SPT with cocycle $\omega$ as defined in Eq.~\ref{eq:omega-def} :
\begin{align}\begin{split}\label{eq:Pure-prediction}
    I_g &= (-1)^{g_a n_a + g_b n_b + g_a g_b n_{ab}} \\
    \implies I_{(1,0)} &= (-1)^{n_a}\\ I_{(0,1)} &= (-1)^{n_b}, \\
    I_{(1,1)} I_{(1,0)}^{-1} I_{(0,1)}^{-1} &= (-1)^{n_{ab}}.
\end{split}\end{align}
$I_g$ is expected to measure the cohomology invariant $\omega(g,g,g)$ \cite{turzillo2025SPT}; when we plug in the explicit cocycle representative in Eq.~\eqref{eq:omegaij-def}, we obtain the right-hand side of Eq.~\eqref{eq:Pure-prediction}.

We now verify Eq.~\eqref{eq:Pure-prediction} for the CZX model with symmetry action given by Eq.~\eqref{eq:PureCZX-SymAction}. We make the simplest choice of $\cD$ that contains a single vertex $o$. In Appendix~\ref{app:origin} we show that our results are the same for plaquette-centered rotations and for larger regions $\cD$. First let us evaluate
\begin{align}\label{eq:Ig2-evaluation}
    I_{g,2} = \operatorname{arg} (\sum_{\{\alpha_l,\beta_l\}} \bra{\beta_{l,p}} U_g^{\cD} C_2^{\cD}  \ket{\alpha_{l,p}} ).
\end{align}
The configurations $\alpha, \beta$ that contribute satisfy $\alpha_{s,l,j} = \beta_{s,l,j}$ whenever $s \neq o$, and $\alpha_{o,l,j} = \beta_{o,l,j+2} + g_l$. Together with the plaquette-GHZ condition, we get $\alpha_{o,l,j} = \alpha_{o,l,j+2} + g_l$. The contribution from a single term $\bra{\beta_{l,p}} U_g^{\cD} C_2^{\cD}  \ket{\alpha_{l,p}}$ subject to the above constraints is determined by the $CZ$ gates at $s = o$ defined in Eq.~\eqref{eq:PureCZX-SymAction}, and equals
\begin{align}\label{eq:Ig-eval}
 (-1)^{( n_a g_a  +  n_{ab}g_{b})\sum_j\alpha_{a,j}\alpha_{a,j+1} +  n_b g_b\sum_j\alpha_{b,j}\alpha_{b,j+1}}.
\end{align}
Note that $\sum_j \alpha_{l,j} \alpha_{l,j+1} = \alpha_{l,1}\alpha_{l,2} + \alpha_{l,2}(\alpha_{l,1}+g_l) + (\alpha_{l,1}+g_l)(\alpha_{l,2}+g_l) + (\alpha_{l,2}+g_l)\alpha_{l,1} = g_l$, which is independent of $\alpha$. Summing over all configurations, we get
\begin{equation}\label{eq:Ig2-result}
    I_{g,2} = (-1)^{g_a n_a + g_b n_b +  g_ag_{b} n_{ab}},
\end{equation}
which is already the right-hand side of Eq.~\eqref{eq:Pure-prediction}. Now let us find $I_{g,4}$. Arguing analogously to Eq.~\eqref{eq:Ig2-evaluation}, two configurations $\alpha, \beta$ that contribute must now satisfy $\alpha_{s,l,j} = \beta_{s,l,j}$ when $s \neq 0$ and $\alpha_{o,l,j} = \beta_{o,l,j+1}+g_l$. This forces $\alpha_{o,l,j}=\alpha_{o,l,l+1}+g_l$, which also means $\alpha_{o,l,j+2}=\alpha_{0,l,j}$. To get $I_{g,4}$ we re-evaluate the right-hand side of Eq.~\eqref{eq:Ig-eval} with these constraints. Now the sum over $j$ is $2\alpha_{l,1}(\alpha_{l,1}+g_l) + 2(\alpha_{l,1}+g_l)= 0 \mod 2$, again independent of $\alpha$. As a result, 
\begin{equation}\label{eq:Ig4-result}
    I_{g,4} = 1 = I_{e,4} \quad \forall g.
\end{equation}
This means there are no crystalline invariants for vertex- or plaquette-centered rotations \cite{turzillo2025SPT}. Combining this with Eq.~\eqref{eq:Ig2-result}, we recover the prediction Eq.~\eqref{eq:Pure-prediction} exactly.

For a finite Abelian group $G = \Z_m \times \Z_n \times \Z_k$, one can also define a `Type-III' cocycle which depends on all three subgroups \cite{propitius1995topological}. The order parameter $I_{g,M}$ cannot detect type-III invariants in general.\footnote{Detecting type-III invariants requires simulating partition functions on $3$-tori in addition to lens spaces \cite{Tantivasadakarn_2017}, and only the latter are related to $I_{g,M}$ \cite{turzillo2025SPT}. In the special case of $G = \Z_2^3$, however, the invariants $I_{g,M}$ happen to suffice to detect the Type-III phase, as the seven nontrivial group elements are enough to determine the seven SPT invariants.} While the CZX model is easily generalized to include Type-III cocycles, this runs into subtleties in redefining $C_M$ so that it commutes with $U_g$ locally \cite{turzillo2025SPT}. Since our main goal is to apply the order parameter to mixed states, we will not focus on Type-III here.  

\textit{Mixed state order parameter.} The CZX models we have defined have the pure density matrix $\rho_0=\bigotimes_p\sum_{\alpha_{l,p},\beta_{l,p}}|\alpha_{l,p}\rangle\langle\beta_{l,p}|$. Let us decohere $\rho_0$ with a dephasing channel $\cE=\otimes_s\cE_s$, where $\cE_s$ is described by Kraus operators $K_0=\sqrt{1-q}\mathds{1}$ and $K_1=\sqrt{q}Z_{s,a,1}$ supported only on the first subsite of $s$ in the $a$ layer, with dephasing probability $q \in [0,1]$. Here we will only discuss the maximally decohered point $q=1/2$, leaving the analysis for general $q$ to Appendix~\ref{app:generalq}. If $p$ is the plaquette containing $(s,1)$, then the condition $\alpha_{a,p} = \beta_{a,p}$ is enforced for each $p$, leading to the simple expression
\begin{equation}\label{eq:rho-1/2}
\rho_{1/2}=\bigotimes_p\sum_{\substack{\alpha_{a,p},\alpha_{b,p},\beta_{b,p}}}|\alpha_{a,p},\alpha_{b,p}\rangle\langle\alpha_{a,p},\beta_{b,p}|~.
\end{equation}
Note that $\rho=\rho_q$ satisfies the strong symmetry condition \cite{Buca_2012,me_open1,ma2023topological,MaWangASPT,lee2023symmetry} $U_g\rho=\rho$ for $g_a = 0$ while it satisfies the weak symmetry condition $U_g\rho U_g^\dagger=\rho$ for all $g$. Therefore the $\Z_2$ subgroup generated by $g_b$ remains strong while the $\Z_2$ subgroup generated by $g_a$ breaks down to a weak symmetry. Importantly, the rotational symmetry also becomes weak.

We need to identify suitable order parameters to detect potential mixed-state SPT invariants in $\rho_q$. It turns out that the simplest generalization of Eq.~\ref{eq:Ig-def}, which is $\operatorname{Tr}(\rho U_g^{\cD} C_M^{\cD})$, does \textit{not} work. We can in fact consider a single CZX layer with a weak $\Z_2$ symmetry (just ignore the $b$ layer in Eq.~\eqref{eq:rho-1/2}). Then vertex-centered rotations give a nonzero expectation value with sign $-1$ when $M=2$, while plaquette-centered rotations identically give zero. We show this in Appendix~\ref{app:origin}. We believe that this order parameter might be failing because it does not account for how the $C_2$ symmetry is weak. This belief motivates us to consider a different quantity
\begin{equation}\label{eq:Jgh-def}
    J_{g,h,M}:=\operatorname{arg}\Tr[U_{gh}^\cD C_M^\cD\sqrt{\rho}C^{\dagger\cD}_M U_{h}^{\dagger\cD}\sqrt{\rho}]
\end{equation}
where we require $g^M = h^M = e$. We demand that $g$ be a strong symmetry, and let $h$ be arbitrary. If we consider the canonical purification of $\rho$, which is the doubled state $\ket{\sqrt\rho}\rangle$ of $\sqrt\rho$ (see \cite{Ma2024SPTdoubled} for a review of doubled-space quantities), then we can write $J$ as
\begin{equation}
    J_{g,h,M} = \langle\bra{\sqrt{\rho}} U_{gh}^\cD C_M^\cD\otimes U_h^\cD C_M^\cD \ket{\sqrt{\rho}}\rangle.
\end{equation}

Defining and characterizing SPT orders in mixed quantum states is a nontrivial task. There are several somewhat different approaches commonly used in the literature. A physically motivated approach is to define SPT phases as equivalence classes of mixed states under two-way connectivity through symmetric finite-depth quantum channels \cite{me_open1,MaWangASPT,ma2023topological}. One can also define mixed-state SPT phases through the pure-state invariants from the doubled Hilbert space (either from the Choi double \cite{Lee2025MixedSPT,Ma2024SPTdoubled,XueLeeBao} or the canonical purification \cite{lu2024bilayer}). These approaches are not obviously equivalent, but they all give the same classification of SPT orders. Here we discuss some relations among these different approaches.

Let us consider a bosonic mixed state $\rho$ with SPT order defined through channel connectivity. For simplicity, we consider the situation with symmetry $G=E\times A$ where $E$ is strong and $A$ is weak, so the SPT orders in low dimensions are classified by $\mathcal{H}^{d+1}(G,U(1))/\mathcal{H}^{d+1}(A,U(1))$ \cite{MaWangASPT,ma2023topological,Lee2025MixedSPT,LessaChengWang,Wang2025AnomalyOQS,Ma2024SPTdoubled}. An important consequence of the two-way channel connectivity is that $\rho$ can be purified to an ordinary SPT state, namely $\rho=\Tr_a|\psi\rangle\langle\psi|$ for some SPT pure state $|\psi\rangle$, where the ancilla $a$ satisfies the usual requirements of being local and trivial under strong symmetries. Furthermore, the topological invariant $\omega\in \mathcal{H}^{d+1}(G,U(1))/\mathcal{H}^{d+1}(A,U(1))$ of $|\psi\rangle$ is purification-independent, as long as the purification is symmetric and short-range entangled. We can therefore use $\omega$ as the topological invariant of the mixed state $\rho$. From this point of view, any (symmetric and short-range entangled) purification gives an equally valid computation of the invariant, and we can just pick the canonical purification $|\sqrt{\rho}\rangle\rangle$, and view the Choi double $|\rho\rangle\rangle$ as a convenient Renyi-$2$ proxy to the canonical purification (as in Appendix \ref{app:generalq}). We note that, however, the exact equivalence between the definitions of SPT through channel connectivity and canonical purification has not been established. A priori, the canonical purification can fail to give the right invariant if $|\sqrt{\rho}\rangle\rangle$ is not short-range entangled, even though $\rho$ is a valid SPT defined through channel connectivity. No such example is known. In fact, existing results appear to support the conjecture that a valid SPT should have a short-range entangled canonical purification.

The canonical purification has an additional advantage compared to other purifications, namely it is sensitive to the so-called strong-to-weak spontaneous symmetry breaking (SWSSB) \cite{lee2023quantum,Lessa2025}. It is known that in the presence of SWSSB, the SPT invariant (such as the topological response) is not well defined \cite{ma2023topological}, and the state $\rho$ is not two-way connected to any SPT state. States with SWSSB can be purified to multiple different SPT phases \cite{Sala2024}, again making the topological invariants ill-defined. The canonical purification $|\sqrt{\rho}\rangle\rangle$ detects SWSSB by hosting ordinary long-range symmetry breaking order in the doubled Hilbert space \cite{Wightman2024,Weinstein2025}. Therefore, if a well-defined topological invariant is extracted from $|\sqrt{\rho}\rangle\rangle$, then at least we are assured that the strong symmetry of the system (which is required to have nontrivial SPT order) is not spontaneously broken. We can generalize this observation by making the following conjecture: if the canonical purification $|\sqrt{\rho}\rangle\rangle$ is a well-defined SPT state, then $\rho$ is a well-defined mixed-state SPT state in the sense of two-way channel connectivity.

\textit{Mixed state CZX calculation.} We are now ready to put the order parameter $J_{g,h,M}$ to the test. Conveniently, our model has $\sqrt{\rho_{1/2}} \propto \rho_{1/2}$ up to an overall positive factor. We have
\begin{align}\label{eq:Jg2-eval}
    &\Tr[U_{gh}^\cD C_M^\cD\sqrt{\rho_{1/2}}C_M^{\dagger \cD}U_h^{\dagger\cD}\sqrt{\rho_{1/2}}]  \propto \nonumber \\  \sum_{\substack{\alpha_a,\alpha_b,\beta_b\\\alpha_a',\alpha_b',\beta_b'}}&\langle\alpha_a',\beta_b'|U_{gh}^\cD C_M^\cD|\alpha_a,\alpha_b\rangle\langle\alpha_a,\beta_b|C_M^{\dagger\cD}U_h^{\dagger\cD}|\alpha_a',\alpha_b'\rangle~.
\end{align}
A nonzero term occurs only when $\alpha_a'=^{U_h^{\cD}C_M^\cD}\alpha_a$ and $\alpha_a'=^{U_{gh}^\cD C_M^{\cD}}\alpha_a$; while $\alpha_b'=^{C_M^\cD}\beta_b$ and $\beta_b'=^{U_{g}^{\cD}C_M^{\cD}}\alpha_b$. Here $^{\mathcal{O}}\alpha_a$ is defined as the classical configuration obtained by applying the operation $\mathcal{O}$ to $\alpha_a$. Below we will abbreviate $^{U_h^{\cD}C_M^\cD}\alpha$ as $^{h,M}\alpha$. In particular, we find $\alpha_a = ^{U_{g}^{\cD}} \alpha_a$, i.e. $g$ does not flip any spins in the $a$ layer. This forces $g_a=0$, i.e. $g$ needs to be a strong symmetry, as we initially assumed. The constraints hold for general $M$ and $\cD$, including both vertex-and plaquette-centered regions, and fix $\alpha_a', \alpha_b'$. Now we evaluate 
\begin{widetext}
\begin{align}\begin{split}
    \label{eq:J-CZX-eval}
J_{g,h,M}&=\arg\Tr[U_{gh}^\cD C_M^\cD\sqrt{\rho_{1/2}}C_M^{\cD\dagger}U_h^{\cD\dagger}\sqrt{\rho_{1/2}}]\\
    &=\arg\sum_{\substack{\alpha_a,\alpha_b,\beta_b\\\alpha_a',\alpha_b',\beta_b'}}\bra{\alpha_a',\beta_b'}U_{gh}^\cD C_M^\cD\ket{\alpha_a,\alpha_b}\bra{\alpha_a,\beta_b}C_M^{\cD\dagger}U_h^{\cD\dagger}\ket{\alpha_a',\alpha_b'}\\
    &=\arg\sum_{\alpha_a,\alpha_b,\beta_b}\bra{{}^{gh,M}\alpha_a,{}^{gh,M}\alpha_b}U_{gh}^\cD C_M^\cD\ket{\alpha_a,\alpha_b}\overline{\bra{{}^{h,M}\alpha_a,{}^{h,M}\beta_b}U_h^\cD C_M^\cD \ket{\alpha_a,\beta_b}}\delta({}^{g}\alpha_a=\alpha_a)\\
    &=\arg\frac{1}{|A|}\sum_{\alpha_a,\alpha_b,\beta_a,\beta_b}\bra{{}^{gh,M}\alpha_a,{}^{gh,M}\alpha_b}U_{gh}^\cD C_M^\cD \ket{\alpha_a,\alpha_b}\overline{\bra{{}^{h,M}\beta_a,{}^{h,M}\beta_b}U_h^\cD C_M^\cD \ket{\beta_a,\beta_b}}\delta({}^{g}\alpha_a=\alpha_a)\\
    &=I_{gh,M}^\text{pure}I_{h,M}^{\text{pure}*}\delta(g_a=0)~.
\end{split}\end{align}
\end{widetext}

To go from the third line to the fourth, we used the fact (demonstrated in Ref. \cite{turzillo2025SPT} for group cohomology models) that the amplitudes $\bra{{}^{g,M}\alpha}U_g^\cD C_M^\cD\ket{\alpha}$ are independent of $\alpha$. The constraint ${}^g\alpha_a=\alpha_a$ means that the component of $g$ in $A$ must vanish, which is to say that $g$ is strong. Using Eq.~\eqref{eq:Pure-prediction}, we find that
\begin{equation}
I_{gh,2}^\text{pure}I_{h,2}^{\text{pure}*} =(-1)^{ g_bn_b + h_ag_bn_{ab}}.
\end{equation}

We can now state our main results. \textit{First}, SPT invariants of the strong symmetry can be detected by setting $h$ to be trivial and varying $g$. \textit{Second}, any putative SPT indices associated only to the weak symmetry (i.e. $g=0$) always cancel out. The index $n_a$ can therefore no longer be detected by $J_{g,h,M}$, and we conclude that $n_a$ is not a well-defined mixed state SPT invariant. \textit{Third}, invariants that depend on both the strong and the weak symmetry can be detected with suitable combinations of $J_{g,h,M}$, in which $g$ and $h$ both need to be non-trivial. Note that $J_{g,h,M}$ can also depend on crystalline topological invariants, but this dependence can be understood from the analysis in Ref.~\cite{turzillo2025SPT}.

\textit{General symmetries.}  Consider a symmetry group $G=E\times A$. The group cohomology models for this symmetry have wavefunction
\begin{equation}
    \ket{\psi_G}=\sum_{\{\alpha_{A,p},\alpha_{E,p}\}}\bigotimes_p\ket{\alpha_{A,p},\alpha_{E,p}}=\sum_{\alpha_A,\alpha_E}\ket{\alpha_A,\alpha_E}~.
\end{equation}
On this pure state, the order parameter for any $g,h$ with $g^M = h^M = e$ evaluates to
\begin{align}\begin{split}
    J_{g,h,M}&=\arg\bra{\psi_G}U_{gh}^\cD C_M^\cD\ket{\psi_G}\bra{\psi_G}C_M^{\cD\dagger}U_h^{\cD\dagger}\ket{\psi_G}\\
    &=I_{gh,M}I_{h,M}^*.
\end{split}\end{align}
Now consider dephasing the $A$ layer to obtain a mixed state
\begin{equation}
    \rho=\sum_{\alpha_A,\alpha_E,\beta_E}\ket{\alpha_A,\alpha_E}\bra{\alpha_A,\beta_E}~,
\end{equation}
where the $A$ symmetry is broken down to a weak symmetry while the $E$ symmetry remains strong. Then we can evaluate the order parameter using the exact same calculation as Eq.~\eqref{eq:J-CZX-eval}, substituting the strong and weak $\Z_2$ symmetries in the CZX example with the general groups $E$ and $A$.

SPT invariants can be extracted from the values of $J_{g,h,M}$ as follows. Pure state SPT phases protected solely by $A$ or $E$ are distinguished by combinations of values $I_{a,M}$ or $I_{b,M}$, respectively, where $a\in A$, $b\in E$. Pure state SPT phases protected jointly by $A$ and $E$ are distinguished by combinations of ratios $I_{ab,M}I_{a,M}^*I_{b,M}^*$. In the mixed state, we can measure
\begin{align}\begin{split}
    I_{b,M}&=J_{b,e,M}~,\\
    I_{ab,M}I_{a,M}^*I_{b,M}^*&=J_{b,a,M}J_{b,e,M}^*~,
\end{split}\end{align}
however $I_{a,M}$ cannot be detected using $J$'s, in agreement with the $\mathcal{H}^3(E \times A,\U(1))/\mathcal{H}^3(A,\U(1))$ classification.

\textit{Measurement protocols.} $I_{g,M}$ and $J_{g,h,M}$ could potentially be measured in quantum simulators. One useful protocol is the \textit{Hadamard test}, which is a standard quantum algorithm to determine the expectation value $\bra{\Psi}U \ket{\Psi}$. We introduce an ancilla in the $\ket{+}$ state, perform a controlled-$U$ operation on the state $\ket{+} \otimes \ket{\Psi}$ and then measure the ancilla qubit in the computational basis. Upon repeating this test, the expectation value of the output turns out to be precisely $\frac{1}{2}\bra{\Psi} (U + U^{
\dagger})\ket{\Psi} = \operatorname{Re}\bra{\Psi}U \ket{\Psi}.$ (On a mixed state $\rho=\sum_ip_i\ket{\psi_i}\bra{\psi_i}$, the Hadamard procedure results in $\operatorname{Re}\Tr(\rho U)=\operatorname{Re}\sum_ip_i\bra{\psi_i}U \ket{\psi_i}$ in expectation.) If we repeat this procedure with the ancilla initialized in the state $\ket{-i} = \frac{1}{\sqrt{2}}(\ket{0}-i\ket{1})$, we will instead compute $\operatorname{Im}\bra{\Psi}U \ket{\Psi}$. In order to measure $I_{g,M}$ we need to choose $U = U_g^{\cD} C_M^{\cD}$. This should be feasible to implement in CZX-type models, where $U_g$ is only composed of $X$ and $CZ$ gates and the region $\cD$ can in theory be small and does not depend on the system size. However, it is not obvious how to measure $J_{g,h,M}$ using the Hadamard test.

Randomized measurements provide a powerful tool to estimate observables that can be expressed as powers of the density matrix. They have previously been used to compute topological invariants in a variety of scenarios \cite{Elben2019RM,RM_Elben20,huang2020RM,RMChern_Cian21,elben2023RMtoolbox}. Below we briefly review the main results needed to measure $I_{g,M}$ and $J_{g,h,M}$. We construct a set of `classical shadows' $\hat{\rho}_k, k = 1,2, \dots, r$ by applying a random single-qubit unitary $U_{j,k}$ to every spin indexed by $j$ in our system, and then measure each qubit in the computational basis. Denote the measurement outcomes by $s_{j,k}$. Then define $\hat{\rho}_k = \bigotimes_j (3U^{\dagger} \ket{s_{j,k}} \bra{s_{j,k}}U - \mathds{1})$. It can be shown that $\mathbb{E}[\hat{\rho}_k] = \rho$ where $\mathbb{E}$ denotes an average over unitaries $U$. For any operators $\hat{O}^{(1)} \in \mathcal{H}$ and $ \hat{O}^{(2)} \in \mathcal{H}^{\otimes 2}$, also define
\begin{align}
    \langle \hat{O}^{(1)} \rangle &:= \frac{1}{r} \sum_{k=1}^r \operatorname{Tr}[\hat{\rho}_k \hat{O}^{(1)}], \\
    \langle \hat{O}^{(2)} \rangle&:= \frac{1}{r(r-1)} \sum_{k_1 \neq k_2} \operatorname{Tr}[\hat{\rho}_{k_1} \hat{\rho}_{k_2} \hat{O}^{(2)}].
\end{align}
It can also be shown that $\mathbb{E}[\langle \hat{O}^{(1)} \rangle] = \operatorname{Tr}[\rho \hat{O}^{(1)}]$, and $\mathbb{E}[\langle \hat{O}^{(2)} \rangle] = \operatorname{Tr}[\hat{O}^{(2)} (\rho \otimes\rho)]$. We can measure $I_{g,M}$ and $J_{g,h,M}$ by setting $\hat{O}^{(1)} = U_g^{\cD} \hat{C}_M^{\cD}$ and $\hat{O}^{(2)} = \mathbb{S} (U_{gh}^{\cD} \hat{C}_M^{\cD} \otimes U_{h}^{\dagger\cD} \hat{C}_M^{\dagger\cD})$, where the operator $\mathbb{S}$ locally swaps the two copies. Obtaining an accuracy of $\epsilon$ in either result is expected to require $r \sim O(2^{c|D|}/\epsilon)$ measurement rounds, where $c$ is some positive number. In experiments that are sufficiently close to the fixed-point limit, $\cD$ needs to contain a very small number of sites, suggesting that the measurement overhead may not be prohibitive.  

\textit{Discussion.} One remaining question is whether the partial symmetry order parameter is robust against the appropriate deformations; in this case, low depth circuits of local channels that satisfy the appropriate weak and strong symmetries. Answering this question appears difficult, however, because it requires a precise understanding of circuits subject to the weak crystalline symmetry. Nevertheless, there is another perspective that suggests the order parameter is topological: recall that it may be interpreted as simulating the partition function of lens spaces \cite{turzillo2025SPT}. By extending this construction to mixed states, the present study opens the door to defining ``partition functions'' of mixed state topological phases (both SPT and intrinsic) in terms of wave functions and their bulk order parameters.

\textit{Acknowledgements.} We thank Tyler Ellison, Matteo Ippoliti and Matteo Votto for discussions, and Jos\'e Garre-Rubio for collaboration on related work. We are grateful to the hospitality of the Erwin Schr\"odinger Institute, where some of this research was performed. Research at Perimeter Institute is supported in part by the Government of Canada through the Department of Innovation, Science and Economic Development and by the Province of Ontario through the Ministry of Colleges and Universities.

\bibliography{ref.bib}
\clearpage
\appendix

\begin{widetext}

\section{Effect of changing the rotation center and the size of $\cD$}\label{app:origin}

\subsection{Pure state CZX calculation}

Here we wish to evaluate $I_{g,2}$ for the bilayer CZX state, allowing the rotation region $\cD$ to be an arbitrarily large square region with either vertex- or plaquette-centered rotations. We wish to show that the value of $I_{g,2}$ does not depend on the choice of rotation center. We follow the computation in Ref.~\cite{turzillo2025SPT}, which applies more generally to group cohomology models on the square lattice. The general expression we need to evaluate is
\begin{align}
    I_{g,2} = \operatorname{arg} (\sum_{\{\alpha_l,\beta_l\}} \bra{\beta_{l,p}} (\hat{U}_g \hat{C}_2)|_D  \ket{\alpha_{l,p}} )
\end{align}
where we have a nonzero matrix element only if $\beta_{l,p} = ^{U_g^{\cD} C_2^{\cD}} \alpha_{l,p}$. In particular, $\beta_{l,p} = \alpha_{l,p}$ whenever the plaquette $p$ contains degrees of freedom outside $\cD$, and $\beta_{l,p} = \alpha_{l,p'} + g_l$ when $p,p' \in \cD$ and are related by a $C_2$ rotation. This further enforces $\alpha_{l,p} = \alpha_{l,p'} + g_l$ when $p$ is a plaquette that straddles $\partial \cD$.

The $CZ$ gates comprising the symmetry action can be grouped into two factors based on whether they act on bulk or boundary degrees of freedom. More precisely, in case (i) the gates act on two degrees of freedom that are separated by a bulk edge  lying completely inside $\cD$, and in case (ii) the degrees of freedom run along the boundary, so there is no such edge.

In the first case, each edge $e$ in $\cD$ lies between two vertices $v_L$ and $v_R$, separating two plaquettes $p_1$ and $p_2$. See Fig.~\ref{fig:CZX-boundary}(a) for an illustration. There are four degrees of freedom $\alpha_{1,L}=\alpha_{1,R}\in p_1, \alpha_{2,L}=\alpha_{2,R} \in p_2$ per layer associated to this edge (we suppress the layer index for the moment). Now the $CZ$ gates associated to the pairs $(\alpha_{1,L},\alpha_{2,L})$ and $(\alpha_{1,R},\alpha_{2,R})$ give equal contributions. All bulk contributions from case (i) can be organized in such pairs and necessarily cancel out, so we are left with the boundary contributions comprising case (ii).

\begin{figure}
    \centering
    \includegraphics[width=0.7\linewidth]{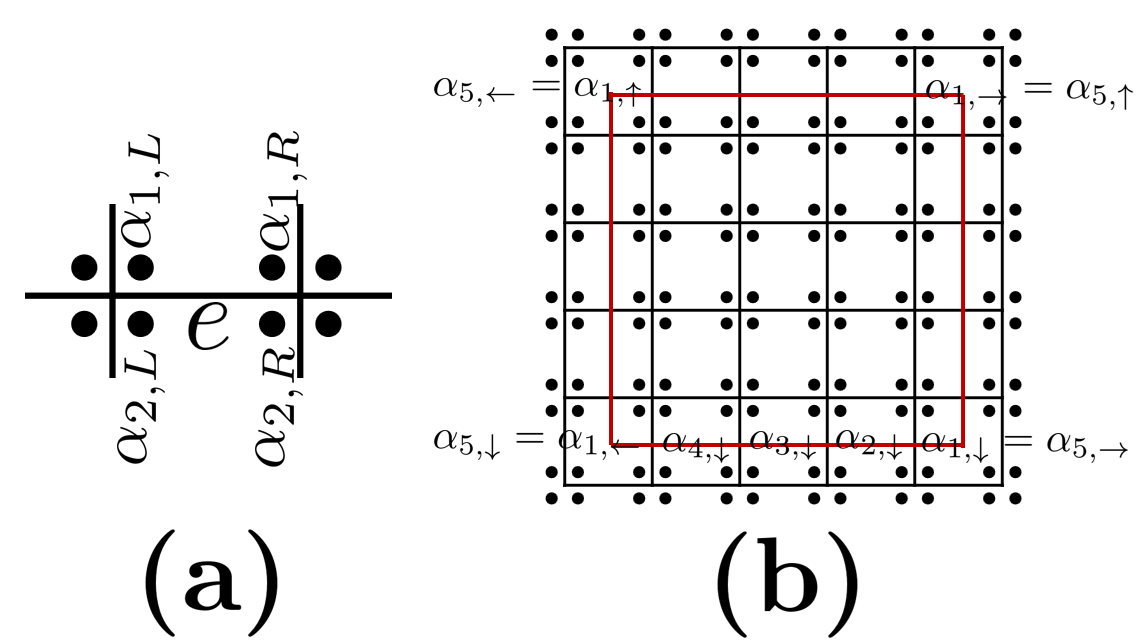}
    \caption{Notation for (a) pairwise cancellation of contributions from $CZ$ gates in the bulk (the case where $e$ is a vertical edge can be handled similarly); (b) boundary plaquette variables, with $L=4$.}
    \label{fig:CZX-boundary}
\end{figure}
Let us label the boundary plaquette variables as $\alpha_{j,*,l}$ where $* = \uparrow, \leftarrow, \downarrow, \rightarrow$ based on the direction of the arrow pointing from the spin to the nearest boundary, see Fig.~\ref{fig:CZX-boundary} (b). The corner variables, which lie equidistant from two boundaries, are assigned $j=1$, and $j$ runs from 1 to $L$ where $L$ is odd for vertex centered rotations, and even for plaquette centered rotations. We use the convention that $\alpha_{L+1,\uparrow,l} \equiv \alpha_{1,\rightarrow,l}$ and so on, as shown in the figure. The boundary condition $\alpha_{l,p} = \alpha_{l,p'} + g_l$ where $p,p'$ are $C_2$ partners translates into the conditions $\alpha_{j,\uparrow,l} = \alpha_{j,\downarrow,l}+g_l$ and $\alpha_{j,\leftarrow,l} = \alpha_{j,\rightarrow,l}+g_l$ (all relations are mod 2). Using these repeatedly, we evaluate the boundary contribution. The total contribution from $CZ$ gates in a single layer $l$ is 
\begin{align}
    & (-1)^{\sum_{j=1}^L \alpha_{j,\uparrow,l}\alpha_{j+1,\uparrow,l} +\alpha_{j,\downarrow,l}\alpha_{j+1,\downarrow,l} +\alpha_{j,\leftarrow,l}\alpha_{j+1,\leftarrow,l} +\alpha_{j,\rightarrow,l}\alpha_{j+1,\rightarrow,l} } \\
    = & (-1)^{\sum_{j=1}^L \alpha_{j,\uparrow,l}\alpha_{j+1,\uparrow,l} + (\alpha_{j,\uparrow,l}+g_l)(\alpha_{j+1,\uparrow,l}+g_l) + \alpha_{j,\leftarrow,l}\alpha_{j+1,\leftarrow,l} + (\alpha_{j,\leftarrow,l}+g_l)(\alpha_{j+1,\leftarrow,l}+g_l)} \\
    = & (-1)^{\sum_{j=1}^L g_l(\alpha_{j,\uparrow,l}+\alpha_{j+1,\uparrow,l} + \alpha_{j,\leftarrow,l}+\alpha_{j+1,\leftarrow,l})} \\
    = & (-1)^{g_l(\cancel{\alpha_{1,\uparrow,l}}+\alpha_{L+1,\uparrow,l}+\alpha_{1,\leftarrow,l}+\cancel{\alpha_{L+1,\leftarrow,l}})} = (-1)^{g_l}.
\end{align}
This result is independent of $L$, i.e. independent of whether the rotation is around a vertex or a plaquette center. It is also independent of the configuration $\alpha$. Combining the contributions from all the $CZ$ gates using Eq.~\eqref{eq:PureCZX-SymAction}, we obtain
\begin{align}
    \bra{\beta_{l,p}} U_g^{\cD} C_2^{\cD}  \ket{\alpha_{l,p}} &=  (-1)^{n_a g_a + n_b g_b + n_{ab} g_a g_b} \times \delta_{\beta_{l,p},^{U_g^{\cD}C_2^{\cD}}\alpha_{l,p}}
\end{align}
and a final sum over $\alpha$ and $\beta$ establishes Eq.~\eqref{eq:Pure-prediction}.
\subsection{Sensitivity of $\operatorname{Tr}(\rho U_g^{\cD} C_M^{\cD})$ to rotation center}

Consider a single layer of the CZX model with decoherence as defined in the main text. Choosing $q=1/2$ results in the mixed state
\begin{equation}
\rho=\bigotimes_p\sum_{\alpha_{a,p}}|\alpha_{a,p}\rangle\langle\alpha_{a,p}|~.
\end{equation}
This mixed state has a single weak $\Z_2$ symmetry with generator $g$. We wish to evaluate $\operatorname{Tr}(\rho U_g^{\cD} C_2^{\cD})$ for both vertex- and plaquette-centered rotations. The results should agree if $\operatorname{Tr}(\rho U_g^{\cD} C_2^{\cD})$ is a reasonable SPT order parameter. Note that 
\begin{align}
    \operatorname{Tr}(\rho U_g^{\cD} C_2^{\cD}) &= \sum_{\alpha_p} \bra{\alpha_p}U_g^{\cD} C_2^{\cD} \ket{\alpha_p}. 
\end{align}
Any configuration state $\ket{\alpha_p}$ which contributes must satisfy $\alpha_p = ^{U_g^{\cD} C_2^{\cD}}\alpha_p$. First consider vertex-centered rotations, where for simplicity $\cD$ encloses a single vertex. Suppose the variables surrounding the central vertex are $\alpha_1, \alpha_2, \alpha_3, \alpha_4$ in clockwise order. The above condition implies that $\alpha_j = \alpha_{j+1} + g \mod 2$. There clearly exist configuration states $\ket{\alpha_p}$ that satisfy this condition, and each of them contributes $(-1)^{\sum_j \alpha_j \alpha_{j+1}} = -1$ to the order parameter, recovering the pure state result. Note that this is already inconsistent with the expectation that decoherence should destroy the SPT invariant of a single $\Z_2$ symmetry.

Next, consider plaquette-centered rotations, where $\cD$ is otherwise arbitrary. Suppose the central plaquette variable in the configuration state $\ket{\alpha_p}$ is $\alpha_0$. The condition for $\ket{\alpha_p}$ to contribute to the order parameter is that $\alpha_0 = \alpha_0 + g$, which forces $g=0$. Here the order parameter is forced to be zero for any nonzero weak symmetry, in contrast to the result for vertex-centered rotations. These inconsistencies mean that we need to consider a different order parameter.

\subsection{Mixed state calculation}

Let us instead consider the order parameter $J_{g,h,M}$ defined in Eq.~\eqref{eq:Jgh2-def}. We use Eq.~\eqref{eq:J-CZX-eval} from the main text:
\begin{align}
    J_{g,h,2}&= \operatorname{arg}\sum_{\substack{\alpha_a,\alpha_b,\beta_b'}}\langle^{U_{gh}C_2^\cD}\alpha_a,\beta_b'|U_{gh}^\cD C_2^\cD|\alpha_a,\alpha_b\rangle(\langle^{U_{gh}C_2^\cD}\alpha_a,\beta_b'|U_h^\cD C_2^\cD|\alpha_a,\alpha_b\rangle)^{\dagger}~
\end{align}
Each term in the above sum is nonzero if and only if $\beta'_b = ^{U_h^{\cD} C_2^{\cD}}\alpha_b$. We can simply recycle the fact derived in the pure state calculation above, that the result is independent of both the origin and the configuration $\alpha$. Any nonzero term has argument $I_{gh,2} I_{h,2}^{-1}$. The technical improvement compared to the unsuccessful attempt in the previous section is that our order parameter is quadratic in $\rho$, meaning we no longer have to take matrix elements between $\alpha_a$ and itself.

    \section{Calculation away from $q=1/2$}\label{app:generalq}

    In this section let us decohere $\rho_0$ with the more general dephasing channel $\cE=\otimes_s\cE_s$, where $\cE_s$ is described by Kraus operators $K_0=\sqrt{1-q}\mathds{1}$ and $K_1=\sqrt{q}Z_{s,a,1}$ supported only on the first subsite of $s$ in the $a$ layer, with dephasing probability $q \in [0,1]$. If $p$ is the plaquette containing $(s,1)$, then $\cE_s$ multiplies the matrix element $\rho_0^{\alpha\beta}$ by the factor
\begin{align}
  \chi^{\alpha\beta}_p(q) &= (1-q + q(-1)^{\alpha_{a,p}+\beta_{a,p}})\nonumber  \\  &= \begin{cases}
      1, & \alpha_{a,p} = \beta_{a,p} \\
      1-2q, & \alpha_{a,p} \neq \beta_{a,p}.
  \end{cases}  
\end{align}
The decohered state is obtained by taking a product over $s$ (which becomes a product over plaquettes $p$), and summing over $\alpha,\beta$:
    \begin{align}
    &\rho_q =\cE(\rho_0) \\ &=\sum_{l,\alpha_{l,p},\beta_{l,p}} \prod_p \chi^{\alpha\beta}_p(q) | \alpha_{a,p},\alpha_{b,p}\rangle\langle\beta_{a,p}\beta_{b,p}|. \label{eq:rho-def}
\end{align}

For the point $q=1/2$, the calculation turns out to be the same whether we compute $J_{g,h,M}$ or $J^{(2)}_{g,h,M}$. For the case with general $q$, it is not easy to work with $\sqrt{\rho_q}$, so we instead work with the R\'enyi 2-version of this quantity:
\begin{equation}\label{eq:Jgh2-def}
    J^{(2)}_{g,h,M}:=\operatorname{arg}\Tr[U_{gh}^\cD C_M^\cD\rho C^{\dagger\cD}_M U_{h}^{\dagger\cD}\rho].
\end{equation}
A similar computation then gives
    \begin{align}\label{eq:J-eval-app}
       \operatorname{Tr}[U^{\cD}_{gh} C_2^{\cD} \rho_q (U_h^{\dagger} C_2^{\dagger})^{\cD} \rho_q]
       &= \sum_{\alpha,\beta,\alpha',\beta'}\prod_{p} \chi^{\alpha'\beta'}_{p}(q)\chi^{\alpha\beta}_{p}(q)\bra{\beta'} U_{gh}^{\cD} C_2^{\cD}\ket{\alpha}\bra{\beta}U_h^{\dagger \cD}C_2^{\cD}\ket{\alpha'} \\
       &\propto \sum_{\alpha,\beta,\alpha',\beta'}\prod_{p} \chi^{\alpha'\beta'}_{p}(q)\chi^{\alpha\beta}_{p}(q) I_{gh,2} I^{-1}_{h,2} \times \delta_{\beta', ^{U_{gh}^{\cD} C_2^{\cD}}\alpha} \delta_{\alpha', ^{U_h^{\cD} C_2^{\cD}}\beta}
    \end{align}
In going from the second line to the third line, we can remove the sum over $\alpha',\beta'$ using the conditions $\beta' = ^{U_{gh}^{\cD} C_2^{\cD}}\alpha$ and $\alpha' = ^{U_h^{\cD} C_2^{\cD}}\beta$. These conditions also enforce that the only configurations which contribute satisfy $\alpha=^{U_{gh}^{\partial \cD} C_2^{\partial \cD}}\alpha,\beta = ^{U_h^{\partial \cD} C_2^{\partial \cD}}\beta$. As a result, the calculation for general $q$ resembles the one for $q=1/2$, under the usual assumption that $g$ is a strong symmetry.

Interestingly, we observe different behaviours as a function of $q$ when we allow $g$ to be a weak symmetry, i.e. $g_a \neq 0$. When $q=1/2$, this immediately forces the trace to vanish, so $J_{g,h,M}$ is no longer well-defined. When $q \neq 1/2$, we find that each summand in Eq.~\eqref{eq:J-eval-app} picks up at least a factor of $(1-2q)$ for each plaquette $p$ in which one of the conditions $\alpha_{a,p} = \alpha_{b,p}$ or $\alpha'_{a,p} = \alpha'_{b,p}$ is violated. When $g_a \neq 0$, these conditions are mathematically inconsistent for each plaquette within $D$, therefore every plaquette in $\cD$ is necessarily violated, and the expectation value is suppressed by a factor of $(1-2q)^{|\cD|}$. Apart from the exponential suppression of the order parameter, when $q>1/2$ each nonzero term can acquire minus signs in a configuration-dependent manner, which means that the phase information is completely lost.

\end{widetext}

\end{document}